\documentclass[aps,prd,floatfix,superscriptaddress,preprintnumbers]{revtex4}
\usepackage{bm,pifont}
\usepackage{amssymb}
\usepackage{amsmath}
\usepackage{amsfonts}
\usepackage{epsfig}
\usepackage{graphicx}
\usepackage{tabularx}
\usepackage{color}
\newcommand{\seq}{\begin{subequations}}
\newcommand{\sen}{\end{subequations}}
\newcommand{\eq}{\begin{eqnarray}}
\newcommand{\en}{\end{eqnarray}}

\def\shiftdown#1{#1\llap{\lower.04ex\hbox{#1}}}

\begin{document}
\title{Updated limits on the CP violating $\eta\pi\pi$ 
and $\eta'\pi\pi$ couplings \\ 
derived from the neutron EDM} 
	
\date{\today}
\author{Alexey S.~Zhevlakov}
\affiliation{
	Department of Physics, Tomsk State University, 634050 Tomsk, Russia} 
\affiliation{Matrosov Institute for System Dynamics and 
	Control Theory SB RAS Lermontov str., 134, 664033, Irkutsk, Russia}
\author{Thomas Gutsche}
\affiliation{Institut f\"ur Theoretische Physik,
	Universit\"at T\"ubingen,
	Kepler Center for Astro and Particle Physics,
	Auf der Morgenstelle 14, D-72076 T\"ubingen, Germany} 
\author{Valery E. Lyubovitskij} 
\affiliation{
	Department of Physics, Tomsk State University, 634050 Tomsk, Russia}
\affiliation{Institut f\"ur Theoretische Physik,
	Universit\"at T\"ubingen,
	Kepler Center for Astro and Particle Physics,
	Auf der Morgenstelle 14, D-72076 T\"ubingen, Germany}
\affiliation{Departamento de F\'\i sica y Centro Cient\'\i fico
	Tecnol\'ogico de Valpara\'\i so-CCTVal, Universidad T\'ecnica
	Federico Santa Mar\'\i a, Casilla 110-V, Valpara\'\i so, Chile}
	
\begin{abstract}
We complete our derivation of upper limits on the CP 
violating $\eta\pi\pi$ and $\eta'\pi\pi$ couplings from an analysis 
of their two-loop contributions to the neutron electric dipole moment (nEDM). 
We use a phenomenological Lagrangian approach which is formulated in terms of 
hadronic degrees of freedom - nucleons 
and pseudoscalar mesons. The essential part of the Lagrangian contains 
the CP violating couplings between $\eta(\eta')$ and pions. 
Previously, we included photons using minimal substitution in case of the 
proton and charged pions. Now we extend our Lagrangian by adding 
the nonminimal couplings, i.e. anomalous magnetic couplings of nucleons 
with the photon. The obtained numerical upper limits for 
the $\eta\pi\pi$ and $\eta'\pi\pi$ couplings 
$|f_{\eta\pi\pi}(M_\eta^2)|<4.4 \times10^{-11}$ and 
$|f_{\eta^\prime\pi\pi}(M_{\eta'}^2)|<3.8 \times10^{-11}$
can be useful for the related, planned experiments at the JLab Eta Factory. 
Using present experimental limits on the nEDM, we derive upper limits on 
the CP violating $\bar\theta$ parameter of 
$\bar{\mathrm{\theta}}<4.7 \times 10^{-10}$. 

\end{abstract}

\maketitle
	
\section{Introduction}
	
Since the 1950s the study of T- or CP- violation in hadronic processes is a 
relevant topic in particle physics since it helps to shed light on 
the entries of the Cabibbo-Kabayashi-Maskawa mixing matrix and the related oscillations
of neutral kaons, $D$ and $B$ mesons. Some phenomena, like 
CP violation in processes involving $K$ and $B$ mesons, 
have been explained in the framework of the Standard Model (SM). 
The study of other CP-violating effects, 
such as strong CP-violation, a neutron electric dipole moment (nEDM), 
decays of $\eta$ and $\eta'$ mesons into two pions, etc. clearly call for 
the search of possible New Physics mechanisms, which are outside the scope of the SM. 
In particular, SM predictions for the nEDM are up to several orders of magnitude lower than 
existing experimental limits. Clearly the study of EDMs of hadrons and  
nuclei could probe New Physics beyond the SM (for a review see, e.g., 
Refs.~\cite{Pospelov:2005pr, Chupp:2017rkp}). Our interest in hadron EDMs 
is motivated by the possibility to extract limits on the CP-violating strong 
coupling between hadrons and the $\bar\theta$ parameter 
(CP-violating gluon-gluon coupling). From study of the nEDM one can estimate the QCD $\bar\theta$ parameter 
and the $\pi NN$, $\eta^{(\prime)} NN$, and $\eta^{(\prime)} \pi\pi$ couplings. 
In series of papers~\cite{Kuckei:2005pg,Dib:2006hk,%
Faessler:2006vi,Faessler:2006at,Castillo-Felisola:2015ema,%
Gutsche:2016jap,Zhevlakov:2018rwo} we gave several analyses
of CP violating physics focused on the nEDM and 
strong CP violating phenomena with the relation to the QCD $\bar\theta$ term, 
aspects of the phenomenology of axions, CP-violating hadronic couplings, 
intrinsic electric and chromoelectric dipole moments of quarks, 
CP-violating quark-gluon, three-gluon and four-quark couplings, etc. 
 
In Refs.~\cite{Gutsche:2016jap,Zhevlakov:2018rwo} we focused on the 
determination of the CP violating couplings $\eta\pi\pi$ 
and $\eta'\pi\pi$ from the nEDM. Our formalism was based on a phenomenological 
Lagrangian describing the interaction of nucleons with pseudoscalar mesons 
(pions and $\eta(\eta')$). The interaction of charged particles with 
photons has been introduced by minimal substitution. In the present paper we extend 
the previous considerations by also including nonminimal couplings induced by the anomalous magnetic moments 
of nucleons. 

\section{Framework}

In this section we review our formalism to link the nEDM to the CP violating couplings.
It is based on a phenomenological 
Lagrangian ${\cal L}_{\rm eff}$ formulated in terms of hadronic degrees of freedom 
(nucleons $N=(p,n)$, pions $\pi=(\pi^\pm,\pi^0)$, $H=(\eta,\eta')$ mesons)  
and photons $A_\mu$ which separates into a free ${\cal L}_{0}$ and 
interaction part ${\cal L}_{\rm int}$ with 
\eq 
{\cal L}_{\rm eff} = {\cal L}_{0} + {\cal L}_{\rm int}\,. 
\en 
${\cal L}_{0}$ includes the usual free terms of nucleons, mesons, and photons 
\eq 
{\cal L}_0 = \bar N (i\not\!\partial - M_N) N
+ \frac{1}{2} \vec{\pi\,} (\Box - M_\pi^2) \vec{\pi\,}
+ \frac{1}{2} H (\Box - M_H^2) H
- \frac{1}{4} F_{\mu\nu} F^{\mu\nu}\,,
\en  
where $\Box = - \partial_\mu \partial^\mu$, 
$F_{\mu\nu} = \partial_\mu A_\nu - \partial_\nu A_\mu$ is the stress 
tensor of the electromagnetic field, $M_N$, $M_\pi$, and $M_H$ are 
the masses of nucleons, pions, and $\eta(\eta')$ mesons, respectively. 
The interaction Lagrangian 
${\cal L}_{\rm int}$ is given by a sum of two part.
The first part contains the strong interaction terms, 
which describe the CP-even couplings of nucleons with pions 
${\cal L}_{\pi NN}$ and $\eta(\eta')$ mesons 
${\cal L}_{HNN}$ and the CP-violating $\eta(\eta')\pi\pi$ 
coupling ${\cal L}^{\rm CP}_{H\pi\pi}$. The second part includes the 
electromagnetic interaction terms, describing the coupling 
of charged pions and nucleons with the photon (${\cal L}_{\gamma NN}$ 
and ${\cal L}_{\gamma\pi\pi}$, respectively): 
\eq\label{Lint} 
{\cal L}_{\rm int} &=& {\cal L}_{\pi NN} + {\cal L}_{\eta(\eta') NN} 
+ {\cal L}^{\rm CP}_{\eta(\eta') \pi\pi} 
+ {\cal L}_{\gamma NN} + {\cal L}_{\gamma\pi\pi}\,, \nonumber\\
{\cal L}_{\pi NN} &=& g_{\pi NN} \bar{N} i\gamma_5 \vec{\pi\,} \vec{\tau\,} N\,,
\nonumber\\
{\cal L}_{HNN} &=& g_{HNN} H \bar{N} i\gamma_5 N\,, \nonumber\\
{\cal L}^{\rm CP}_{H\pi\pi} &=& f_{H\pi\pi} M_H H \vec{\pi\,}^2 \,,\nonumber\\
{\cal L}_{\gamma NN} &=& e A_\mu \bar{N} \Big(\gamma^\mu Q_N 
+ \frac{i \sigma^{\mu\nu} q_\nu}{2 M_N} k_N \Big) N\,, \nonumber\\
{\cal L}_{\gamma\pi\pi} &=& e A_\mu \Big(\pi^- i\partial^\mu \pi^+ 
- \pi^+ i\partial^\mu \pi^-\Big) 
+ e^2 A_\mu A^\mu \pi^+ \pi^- \,, 
\en  
where $g_{\pi NN} = \frac{g_A}{F_\pi}M_N$, 
$g_A = 1.275$ is the nucleon axial charge, 
$F_\pi = 92.4$ MeV is the pion decay constant,  
$g_{HNN}$ and $f_{H\pi\pi}$ are corresponding CP-even and 
CP-odd couplings between pions and $\eta(\eta')$,  
$\gamma^\mu$ and $\gamma^5$ are the Dirac matrices,  
and $\sigma^{\mu\nu} = \frac{i}{2}[\gamma^\mu,\gamma^\nu]$. 
The values of $g_{\eta NN}$ and $g_{\eta^\prime NN}$ are taken 
from Ref.~\cite{Tiator:2018heh}: 
$g_{\eta NN} = g_{\eta^\prime NN} = 0.9$. 
Note that in the case of nucleons we include both minimal and nonminimal electromagnetic couplings. 
Here $Q_N = {\rm diag}(1,0)$ and $k_N = {\rm diag}(k_p,k_n)$ 
are the diagonal matrices of nucleon charges and anomalous magnetic 
moments, respectively, where $k_p = 1.793$ and $k_n = - 1.913$. 
For the  CP-even interactions of 
nucleons with pseudoscalar mesons we use the pseudoscalar (PS) 
coupling~\cite{Weinberg:1968de}, which is equivalent 
to the pseudovector (PV) coupling as demonstrated in 
Refs.~\cite{Weinberg:1968de,Faessler:2005gd,Lensky:2009uv}. 
As we have shown in Ref.~\cite{Zhevlakov:2018rwo}, matrix elements in the two theories 
can differ by a divergent term, which can always be absorbed by 
an appropriative choice of a counterterm. In particular, in the case of the matrix element 
describing the nEDM the PS theory does not contain a logarithmic divergence, 
while it occurs in the PV theory. 

The CP-violating term ${\cal L}^{\rm CP}_{H\pi\pi}$ 
induces a contribution to the nEDM. The $\eta(\eta')\pi\pi$ couplings 
define the corresponding two-body decay branching ratios as 
\begin{align}
{\rm Br}(H\to\pi\pi) = 
n_\Gamma \, \frac{\sqrt{M_H^2-4M_\pi^2}}{4\pi\Gamma_H^{tot}}f_{H\pi\pi}^2\,,
\label{eq:etapipi_decay_coupling}
\end{align}
where $\Gamma_H^{tot}$ is the total width of $H$; 
$n_\Gamma$ is a final-state factor, which equals 1/2 
for the $\pi^0\pi^0$ and 1 for the $\pi^+\pi^-$ final states. 
Upper limits for these decays are set 
by the LHCb results~\cite{Aaij:2016jaa}  
	\begin{align}
	\rm{Br}(\eta(\eta')\to\pi\pi) &<
	\left\{
	\begin{array}{l}
	1.3 (1.8)\times 10^{-5},\;\pi^+\pi^-\\
	3.5 (4.0)\times 10^{-4},\;\pi^0\pi^0\\
	\end{array}
	\right. \, .
	\label{eq:explimits}
	\end{align}

As discussed in Ref.~\cite{Zhevlakov:2018rwo}, 
there are two possible mechanisms for the generation of the $\eta(\eta')\pi\pi$ 
effective couplings. In the first mechanism this coupling 
is generated by the QCD $\bar\theta$-term~\cite{Crewther:1979pi,ShifmanB166}
\eq\label{QCD_fetapipi}
f_{\eta\pi\pi}^{\bar\theta} = -\frac{1}{\sqrt{3}} \,
\frac{\bar\theta\, M_\pi^2\, R}{F_\pi\, M_\eta \, (1+R)^2}\,, \quad 
f_{\eta'\pi\pi}^{\bar\theta} = \sqrt{2} \, 
f_{\eta\pi\pi}^{\bar\theta} \, \frac{M_\eta}{M_{\eta'}}\,, 
\en
where $\bar\theta$ is the QCD vacuum angle and $R = m_u/m_d$ is the
ratio of $u$ and $d$ current quark masses. 
In this scenario, the $\eta(\eta')\pi\pi$ couplings 
are proportional to $\bar\theta$, which in turn is originally constrained by the 
experimental bounds on the neutron EDM~\cite{Harris:1999jx,Baker:2006ts}. 
In the the second scenario the nEDM and the CP violating 
$\eta\to\pi\pi$ vertices are generated by two distinct mechanisms, 
without specifying details of a particular model in which this scenario 
would be realized. Thereby one can expect that the yet unknown mechanisms 
due to New Physics could enlarge the $\eta(\eta')\pi\pi$ 
couplings, which would induce a contribution to the nEDM at the two-loop level 
(see details in Ref.~\cite{Zhevlakov:2018rwo}). 

\section{Neutron EDM induced by the CP violating $\eta(\eta')\pi\pi$ couplings 
at the two-loop level}

In our approach a neutron EDM is described by a set of two-loop diagrams 
shown in Fig.~\ref{two_loop_1} and~\ref{two_loop}.
In Ref.~\cite{Zhevlakov:2018rwo} we already evaluated the diagrams generated by the 
minimal electromagnetic coupling [see Fig.\ref{two_loop_1}], 
i.e., by the coupling of virtual charged pions and the proton to the electromagnetic field. 
Here we extend our analysis by inclusion of the nonminimal couplings of the nucleon
to the electromagnetic field due to the anomalous magnetic moments $k_N$ [see Fig.~\ref{two_loop}].
contained in the interaction Lagrangian ${\cal L}_{\gamma NN}$.

	\begingroup
	\begin{figure}		
\includegraphics[width=0.99\textwidth,trim={2cm 22.5cm 0cm 2cm},clip]{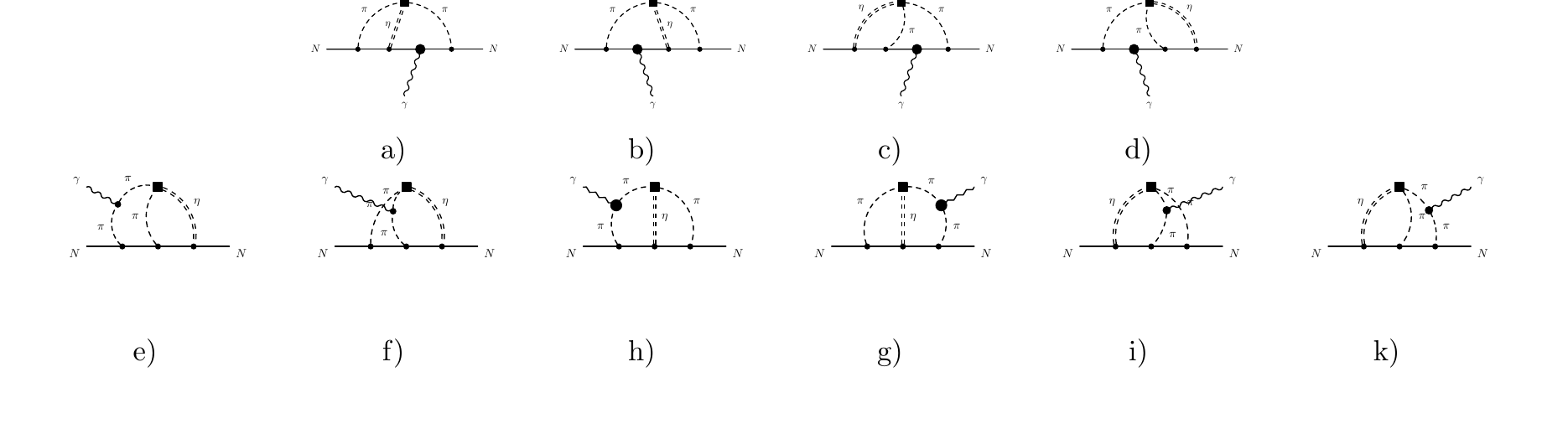}			
		\caption{Diagrams contributing to the nEDM which are induced by the 
			minimal electromagnetic couplings of proton and charged pions. 
			The solid square denotes the CP-violating $\eta\,\pi^+\pi^-$ vertex.}
        \label{two_loop_1}
	\end{figure}
	\endgroup	
	
	\begingroup
	\begin{figure}	[t]	
\includegraphics[width=0.99\textwidth,trim={2cm 22.5cm 2cm 2cm},clip]{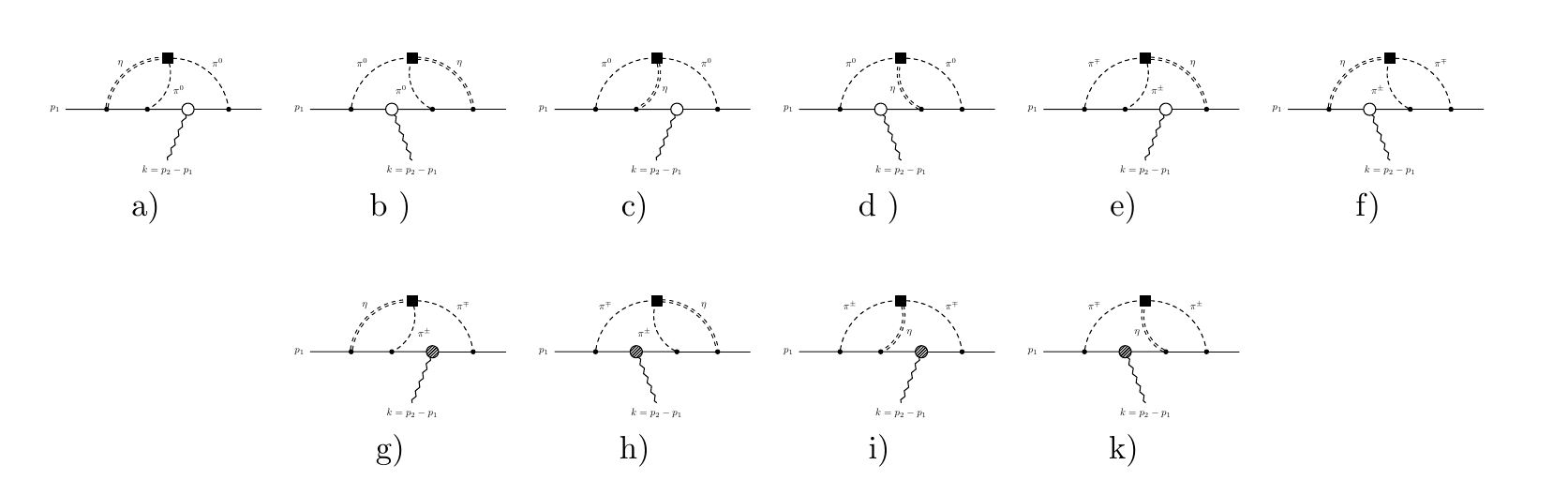}			
\caption{Diagrams contributing to the nEDM which are induced by the
nonminimal electromagnetic couplings (anomalous magnetic moments) of nucleons. 
The solid square denotes the CP-violating $\eta(\eta^\prime)\,\pi\pi$ vertex. 
Empty and shaded circles correspond to the nonminimal electromagnetic 
couplings of neutron and proton, respectively.}\label{two_loop}
\end{figure}
\endgroup

The matrix element corresponding to the diagrams of Figs.~\ref{two_loop_1} 
and~\ref{two_loop} or the electromagnetic vertex function of the neutron 
is expanded in terms of four relativistic form factors $F_E$ (electric), 
$F_M$ (magnetic), $F_D$ (electric dipole) and $F_A$ (anapole) as 
\eq
M_{\rm inv}&=&\bar u_N(p_2)\,\Gamma^\mu(p_1,p_2)\,u_N(p_1)\,, 
\quad\quad \Gamma^\mu(p_1,p_2) \,=\, 
\gamma^{\mu} \, F_E(q^2) 
\,+\,\frac{i}{2 M_N} \sigma^{\mu\nu}q_{\nu} \, F_M(q^2)\nonumber\\
&+&\frac{1}{2 M_N}\sigma^{\mu\nu}q_{\nu} \gamma^5 \, F_D(q^2)
\,+\, \frac{1}{M_N^2} (\gamma^\mu q^2 - 2M_N q^{\mu}) \gamma^5
\, F_A(q^2)
\label{vertex}
\en
where $p_1$ and $p_2$ are the momenta of the initial and final neutron states, 
$q^2=(p_2-p_1)^2$ is the transfer momentum squared. 
The nEDM is defined as $d_n^E=-F_D(0)/(2 M_N)$. 

To proceed we have to evaluate two-loop diagrams in the framework of the PS approach for
the coupling between nucleons and pseudoscalar mesons. 
We want to point out again that the diagrams generated by the minimal coupling 
of charged pions and the proton with the electromagnetic field 
have been calculated in Ref.~\cite{Zhevlakov:2018rwo}. 
Also in Ref.~\cite{Zhevlakov:2018rwo} one can find details of 
the calculational technique,  which is the same for the diagrams 
in Fig.~\ref{two_loop} involving the nonminimal electromagnetic 
couplings of nucleons (anomalous magnetic moments).
The diagrams in Fig.~\ref{two_loop} can be grouped into sets 
with the same topology: (2a and 2b), (2c and 2d), (2e and 2f), 
(2g and 2h), and (2i and 2k).   

The generic contribution of these diagrams to the nEDM is written as: 
\eq 
- \bar u_N(p_2) \, d^i_N \, \sigma^{\mu\nu}q_{\nu} \gamma^5\,  u_N(p_1) 
+ \ldots 
&=& f_{H\pi\pi} \, g_{HNN} \, g_{\pi NN}^2 \, M_H \, \frac{k_N}{2 M_N} 
\ I_{\rm loop}\, .
\en 
$I_{\rm loop}$ is sum of two topologically equivalent Feynman diagrams
\eq 
I_{\rm loop} &=& 
\int \frac{d^4q_1 d^4 q_2}{(2\pi)^8} \, 
S_{M_1}(q_1) S_{M_1}(q_2) S_{M_3}(q_2-q_1) \, 
 \nonumber\\[2mm]
&\times& \left[\bar u_N(p_2) \left( \, \gamma_5 S_N(p_2 + q_2) \, 
\sigma^{\mu\nu} q_{\nu} \, 
S_N(p_1 + q_2) \, \gamma_5 \, S_N(p_1+q_1) \, \gamma_5  \right.\right. 
\nonumber \\[2mm]
&+& \gamma_5 S_N(p_2 + q_2) \, \gamma_5
S_N(p_2 + q_1) \, \sigma^{\mu\nu} \, q_{\nu} \, S_N(p_1+q_1) \, \gamma_5
 \left.\left.\right) 
 u_N(p_1)\right] 
\nonumber\\[2mm]	
&=& - \bar u_N(p_2)  \, \sigma^{\mu\nu} q_{\nu} \, \gamma^5 \,  u_N(p_1) \, 
I_{d}(M_1,M_2,M_3)  \,, 
\nonumber\\[2mm]
I_{d}(M_1,M_2,M_3) 
&=& \frac{2M_N}{(4\pi)^4} \, 
\int\limits_0^1 d\alpha_1 \cdots \int\limits_0^1 d\alpha_6 \,
\frac{\delta\Big(1-\sum\limits_{i=1}^6\alpha_i\Big)} 
{\Delta + B A^{-1} B} \, 
\biggl[ 
-1 + 3 A_{12}^{-1} \beta_2 + \frac{M_N^2\beta_1\beta_2}{\Delta + B A^{-1} B} \, 
\, (2 - \beta_2) \biggr] \,, 
\en
where $S_N(k) = (\not\!k - M_N)^{-1}$ and $S_{M_i}(k) = (k^2 - M_i^2)^{-1}$ 
are the nucleon and meson (with mass $M_i$) propagators, respectively. 
Here $A_{ij}$ is the $2\times 2$ matrix
\eq 
A_{ij} &=& \left(
\begin{array}{cc}
\alpha_{146} & - \alpha_6   \\
- \alpha_6    & \alpha_{2356} \\
\end{array}
\right)\,, \quad \alpha_{i_1\cdots i_k} =
\alpha_{i_1} + \ldots + \alpha_{i_k}, \,
\en 
$A^{-1}$ and ${\rm det} A$ are its inverse 
and determinant, respectively, and
$B_1 = p_1 \alpha_1\,,  \                                       
 B_2 \,=\, p_1 \alpha_3 + p_2 \alpha_2\,, \                             
\Delta = M_1^2 \alpha_4 + M_2^2 \alpha_5 + M_3^2 \alpha_6$, 
$\beta_1=\alpha_1A_{11}^{-1}+\alpha_{23}A_{12}^{-1}$ and  
$\beta_2=\alpha_1A_{12}^{-1}+\alpha_{23}A_{22}^{-1}$.  
$I_{d}(M_1,M_2,M_3)$ is the scalar function deduced after 
calculation of the generic two-loop diagram from the set in Fig.~\ref{two_loop}.  

Summing all graphs of Fig.~\ref{two_loop} we obtain the resulting  
expression for the nEDM 
\eq 
d_N &=& 2 M_\eta f_{\eta\pi\pi} g_{NN\eta} g_{NN\pi}^2 \,  
\biggl[ k_n \Big( I_{d}(M_\pi,M_\pi,M_\eta)
+\frac{1}{2} I_{d}(M_\pi,M_\eta,M_\pi)+\frac{1}{2} I_{d}(M_\eta,M_\pi,M_\pi) 
\Big)+ \\
&+& 
k_p \Big(I_{d}(M_\pi,M_\eta,M_\pi)+I_{d}(M_\eta,M_\pi,M_\pi)\Big) \biggr]\,. 
\nonumber
\en 
Here the factor $1/2$ corresponds to the graphs with neutral pion loops. 

\section{Results and discussion}	

Our numerical result for the nEDM induced by the CP violating $\eta(\eta')\pi\pi$ 
couplings and the anomalous magnetic moments of nucleons is 
\eq 
& &d^{E,k}_n\simeq (c_\eta f_{\eta\pi\pi}
+ c_{\eta'} f_{\eta^\prime\pi\pi}) \times 10^{-16} \, \text{e} \cdot \text{cm}\,, 
\nonumber\\
& &c_\eta = -0.14\,, \quad 
c_{\eta'} = -0.22\,. 
\en
The full result including both minimal and nonminimal electromagnetic couplings 
of nucleons can easily be computed by taking into account our previous results 
of Ref.~\cite{Zhevlakov:2018rwo} restricted to the case of minimal coupling: 
\eq\label{extract_etapipi} 
& &d^{E}_n\simeq(c_\eta f_{\eta\pi\pi}
+ c_{\eta'} f_{\eta^\prime\pi\pi})\times 10^{-16} \, \text{e} \cdot \text{cm}\,, 
\nonumber\\
& &c_\eta = 6.62\,, \quad 
c_{\eta'} = 7.64\,. 
\en
In Table~\ref{tab:num_nEDM} we present the detailed numerical results for 
the contribution of each diagram and their total contribution to 
the couplings $c_\eta$ and $c_{\eta'}$.
For each diagram we specify (if it occurs) the contribution 
of charged pion-photon (PP) coupling, 
nucleon-photon minimal (MC) and nonminimal (NC) couplings 
and also indicate their total contribution (PP+MC+NC). 
The contributions coming from the nonminimal coupling of proton and neutron 
to the electromagnetic field have the same order of magnitude as the one 
induced by minimal coupling of the proton, but they compensate 
each other due to their oposite sign. The total numerical contribution of the nonminimal 
couplings of the nucleon is relatively suppressed (by one order of magnitude) 
compared to the total contribution of the minimal coupling of the proton. 

\begin{table}[ht]
\begin{center}
\caption{Numerical results for the $c_\eta$ and $c_{\eta'}$ couplings.} 
\label{tab:num_nEDM}
\def\arraystretch{1.25}
\begin{tabular}{|c|c|c|c|c|c|c|c|c|}
\hline
Diagram & \multicolumn{4}{c|}{Coupling $c_\eta$} 
        & \multicolumn{4}{c|}{Coupling $c_{\eta'}$}       \\
\cline{2-9}
(Figs.\ref{two_loop_1} and \ref{two_loop})       
& PP & MC & NC & PP+MC+NC & PP & MC & NC & PP+MC+NC \\
\cline{1-9}
1a + 1b / 2i+2k &- &0.58 &-0.92 &-0.34&- &0.71 &-1.57 & -0.86\\
\hline 
1c + 1d / 2g + 2h &- &0.56 & -0.88 &-0.32 &- &0.67 & -1.5 &-0.83\\
\hline 
1e + 1k  &1.12 &-& -&1.12 &1.29 &-&- &1.29\\
\hline 
1g + 1h  &1.02 &- & -&1.02 & 1.13 &- &- &1.01\\
\hline
1f + 1i  &0.1 &- & -&0.1 &0.13 &-  &- &0.13\\
\hline  
2a + 2b  &- &- & 0.77& 0.77 &- &- & 1.32& 1.32\\
\hline
2c + 2d  &- &- & 0.47& 0.47 &- &- & 0.8& 0.8\\
\hline
2e + 2f  &- &- & 0.49 & 0.49&- &- & 0.84 & 0.84\\
\hline
Total   & 4.48& 2.28&-0.14 &6.62 & 5.1 &2.76 &-0.22 & 7.64 \\ 
\hline
\end{tabular}
\end{center}
\end{table}

The bounds for the branching ratios of the rare decays  $\Gamma_{\eta\pi\pi}$ 
and $\Gamma_{\eta^\prime\pi\pi}$ are strongly suppressed 
when compared to existing data~\cite{Aaij:2016jaa} 
\begin{align}
&\mathrm{Br}(\eta\to\pi^+\pi^-)<5.54 \times10^{-17}\,, 
&\mathrm{Br}(\eta^\prime\to\pi^+\pi^-)<5.33\times10^{-19}\,,
\\
&
\mathrm{Br}(\eta\to\pi^0\pi^0)<2.27 \times10^{-17}\,, 
&\mathrm{Br}(\eta^\prime\to\pi^0\pi^0)<2.17\times10^{-19}\,.
\label{G_etapp}
\end{align}
When we deduce these new bounds we suppose that the CP-violating $\eta\pi\pi$ and $\eta^\prime\pi\pi$ 
couplings are independent and use the current experimental bound on the nEDM: 
$|d_n^E|<2.9 \times 10^{-26}  \,\text{e} \cdot \text{cm}$.   
These limits are about $\sim$ 12-14 orders of magnitude more stringent than given by the recent data from the 
LHCb Collaboration~\cite{Aaij:2016jaa}. 
The planned study of the $\eta(\eta')\pi\pi$ decays at the JLab Eta Factory 
(JEF)~\cite{Gan:2017kfr} could shed light on the possible impact of New Physics 
on these CP-violating processes. 

The CP-violating  $\eta\pi\pi$ and $\eta^\prime\pi\pi$ couplings
are estimated using Eq.~(\ref{extract_etapipi}) 
and limits on the nEDM~\cite{Tanabashi:2018oca}:  
\begin{equation}
|f_{\eta\pi\pi}(m_\eta^2)|<4.4 \times10^{-11},\qquad\qquad
|f_{\eta^\prime\pi\pi}(m_{\eta'}^2)|<3.8 \times10^{-11}\,.
\label{f_etapp}
\end{equation}
Note that these results are very close to the ones obtained 
in Ref.~\cite{Zhevlakov:2018rwo} restricted to 
the minimal coupling of charged pions and proton with the photon: 
\eq 
|f_{\eta\pi\pi}(m_\eta^2)|<4.3 \times10^{-11},\qquad\qquad
|f_{\eta^\prime\pi\pi}(m_{\eta'}^2)|<3.7 \times10^{-11}\,.
\label{f_etapp_NM}
\en  
Using Eq.~(\ref{QCD_fetapipi}) we also derive upper limits 
for the $\bar\theta$ parameter: 
\eq 
\bar\theta^\eta    <  9.1 \times 10^{-10}\,, \quad 
\bar\theta^{\eta'} <  9.7\times 10^{-10}\,, \quad    
\en 
for a current quark mass ratio $R=0.556$ taken from ChPT at 1 GeV scale~\cite{Gasser:1982ap} 
and 
\eq 
\bar\theta^\eta    <  8.6\times 10^{-10}\,, \quad 
\bar\theta^{\eta'} <  9.1\times 10^{-10}\,, \quad    
\en 
for $R=0.468$ taken from QCD lattice (LQCD) data at scale of 2 GeV~\cite{Tanabashi:2018oca}.

One can also pursue another way to obtain an upper estimate for the $\bar\theta$ parameter. 
In particular, one can extract $\bar\theta$ substituting the QCD relations between 
$f^{\bar\theta}_{\eta(\eta')\pi\pi}$ and $\bar\theta$ (\ref{QCD_fetapipi}) into 
the expression for the nEDM~(\ref{extract_etapipi}). 
For the quark mass ratios taken from ChPT and LQCD we get: 
\eq 
& &d^E_n \simeq (c^E_n \times 10^{-16}) \cdot \bar\theta \
\text{e} \cdot \text{cm}\,, \nonumber\\
& &c^E_n = 0.65 \ ({\rm ChPT})\,, \quad 
   c^E_n = 0.62 \ ({\rm LQCD})\,.
\en 
Using data on the nEDM~\cite{Graner:2016ses} we extract the following upper limits for $\bar\theta$ 
\eq 
\bar\theta = 4.4 \times 10^{-10} \ ({\rm ChPT})\,, \quad 
\bar\theta = 4.7 \times 10^{-10} \ ({\rm LQCD})\,. 
\en 
The second predictions of upper limits for $\bar\theta$ are by a factor 2 smaller than the first ones and 
are closer to the prediction done in Ref.~\cite{Sahoo:2016zvr}. In particular, in Ref.~\cite{Sahoo:2016zvr} 
the value  $\bar\theta  \sim 1.1 \times 10^{-10}$ was extracted using data for the nEDM with 
$|d_n^E|<1.6 \times 10^{-26} \,\text{e} \cdot \text{cm}$. For this limit on the nEDM we deduce  
$\bar\theta = 2.4 \times 10^{-10} \ ({\rm ChPT})$ and 
$\bar\theta = 2.6 \times 10^{-10} \ ({\rm LQCD})$.  

In conclusion, we studied limits on the QCD CP- violating parameter $\bar\theta$ 
and branchings of the CP- violating rare decays  $\eta \to \pi\pi$ and 
$\eta^\prime \to \pi\pi$ using a phenomenological Lagrangian approach.
We particularly took into account both minimal and nonminimal couplings of the nucleon to the photon. 
The nEDM was induced by the CP violating $\eta(\eta') \to \pi\pi$ couplings. 
Obtained results will be important for the planned experiments on rare 
$\eta$ and $\eta'$ meson decays at JEF~\cite{Gan:2017kfr}. 

\begin{acknowledgments} 

The work of A.S.Zh. was funded by Russian Science Foundation grant 
(RSCF 18-72-00046), in particular part of numerical calculation.  
The work of V.E.L. was funded by the Carl Zeiss Foundation under 
Project ``Kepler Center f\"ur Astro- und
Teilchenphysik: Hochsensitive Nachweistechnik zur Erforschung des
unsichtbaren Universums (Gz: 0653-2.8/581/2)'', and 
by CONICYT (Chile) under Grant No. PIA/Basal FB0821. 

\end{acknowledgments}


\begin{thebibliography}{23}
	\expandafter\ifx\csname natexlab\endcsname\relax\def\natexlab#1{#1}\fi
	\expandafter\ifx\csname bibnamefont\endcsname\relax
	\def\bibnamefont#1{#1}\fi
	\expandafter\ifx\csname bibfnamefont\endcsname\relax
	\def\bibfnamefont#1{#1}\fi
	\expandafter\ifx\csname citenamefont\endcsname\relax
	\def\citenamefont#1{#1}\fi
	\expandafter\ifx\csname url\endcsname\relax
	\def\url#1{\texttt{#1}}\fi
	\expandafter\ifx\csname urlprefix\endcsname\relax\def\urlprefix{URL }\fi
	\providecommand{\bibinfo}[2]{#2}
	\providecommand{\eprint}[2][]{\url{#2}}
	
	\bibitem[{\citenamefont{Pospelov and Ritz}(2005)}]{Pospelov:2005pr}
	\bibinfo{author}{\bibfnamefont{M.}~\bibnamefont{Pospelov}} \bibnamefont{and}
	\bibinfo{author}{\bibfnamefont{A.}~\bibnamefont{Ritz}},
	\bibinfo{journal}{Annals Phys.} \textbf{\bibinfo{volume}{318}},
	\bibinfo{pages}{119} (\bibinfo{year}{2005}).
	
	\bibitem[{\citenamefont{Chupp et~al.}(2019)\citenamefont{Chupp, Fierlinger,
			Ramsey-Musolf, and Singh}}]{Chupp:2017rkp}
	\bibinfo{author}{\bibfnamefont{T.}~\bibnamefont{Chupp}},
	\bibinfo{author}{\bibfnamefont{P.}~\bibnamefont{Fierlinger}},
	\bibinfo{author}{\bibfnamefont{M.}~\bibnamefont{Ramsey-Musolf}},
	\bibnamefont{and} \bibinfo{author}{\bibfnamefont{J.}~\bibnamefont{Singh}},
	\bibinfo{journal}{Rev. Mod. Phys.} \textbf{\bibinfo{volume}{91}},
	\bibinfo{pages}{015001} (\bibinfo{year}{2019}).
	
	\bibitem[{\citenamefont{Kuckei et~al.}(2007)\citenamefont{Kuckei, Dib,
			Faessler, Gutsche, Kovalenko, Lyubovitskij, and Pumsa-ard}}]{Kuckei:2005pg}
	\bibinfo{author}{\bibfnamefont{J.}~\bibnamefont{Kuckei}},
	\bibinfo{author}{\bibfnamefont{C.}~\bibnamefont{Dib}},
	\bibinfo{author}{\bibfnamefont{A.}~\bibnamefont{Faessler}},
	\bibinfo{author}{\bibfnamefont{T.}~\bibnamefont{Gutsche}},
	\bibinfo{author}{\bibfnamefont{S.}~\bibnamefont{Kovalenko}},
	\bibinfo{author}{\bibfnamefont{V.~E.} \bibnamefont{Lyubovitskij}},
	\bibnamefont{and}
	\bibinfo{author}{\bibfnamefont{K.}~\bibnamefont{Pumsa-ard}},
	\bibinfo{journal}{Phys. Atom. Nucl.} \textbf{\bibinfo{volume}{70}},
	\bibinfo{pages}{349} (\bibinfo{year}{2007}).
	
	\bibitem[{\citenamefont{Dib et~al.}(2006)\citenamefont{Dib, Faessler, Gutsche,
			Kovalenko, Kuckei, Lyubovitskij, and Pumsa-ard}}]{Dib:2006hk}
	\bibinfo{author}{\bibfnamefont{C.}~\bibnamefont{Dib}},
	\bibinfo{author}{\bibfnamefont{A.}~\bibnamefont{Faessler}},
	\bibinfo{author}{\bibfnamefont{T.}~\bibnamefont{Gutsche}},
	\bibinfo{author}{\bibfnamefont{S.}~\bibnamefont{Kovalenko}},
	\bibinfo{author}{\bibfnamefont{J.}~\bibnamefont{Kuckei}},
	\bibinfo{author}{\bibfnamefont{V.~E.} \bibnamefont{Lyubovitskij}},
	\bibnamefont{and}
	\bibinfo{author}{\bibfnamefont{K.}~\bibnamefont{Pumsa-ard}},
	\bibinfo{journal}{J. Phys.} \textbf{\bibinfo{volume}{G32}},
	\bibinfo{pages}{547} (\bibinfo{year}{2006}).
	
	\bibitem[{\citenamefont{Faessler
			et~al.}(2006{\natexlab{a}})\citenamefont{Faessler, Gutsche, Kovalenko, and
			Lyubovitskij}}]{Faessler:2006vi}
	\bibinfo{author}{\bibfnamefont{A.}~\bibnamefont{Faessler}},
	\bibinfo{author}{\bibfnamefont{T.}~\bibnamefont{Gutsche}},
	\bibinfo{author}{\bibfnamefont{S.}~\bibnamefont{Kovalenko}},
	\bibnamefont{and} \bibinfo{author}{\bibfnamefont{V.~E.}
		\bibnamefont{Lyubovitskij}}, \bibinfo{journal}{Phys. Rev.}
	\textbf{\bibinfo{volume}{D73}}, \bibinfo{pages}{114023}
	(\bibinfo{year}{2006}{\natexlab{a}}).
	
	\bibitem[{\citenamefont{Faessler
			et~al.}(2006{\natexlab{b}})\citenamefont{Faessler, Gutsche, Kovalenko, and
			Lyubovitskij}}]{Faessler:2006at}
	\bibinfo{author}{\bibfnamefont{A.}~\bibnamefont{Faessler}},
	\bibinfo{author}{\bibfnamefont{T.}~\bibnamefont{Gutsche}},
	\bibinfo{author}{\bibfnamefont{S.}~\bibnamefont{Kovalenko}},
	\bibnamefont{and} \bibinfo{author}{\bibfnamefont{V.~E.}
		\bibnamefont{Lyubovitskij}}, \bibinfo{journal}{Phys. Rev.}
	\textbf{\bibinfo{volume}{D74}}, \bibinfo{pages}{074013}
	(\bibinfo{year}{2006}{\natexlab{b}}).
	
	\bibitem[{\citenamefont{Castillo-Felisola
			et~al.}(2015)\citenamefont{Castillo-Felisola, Corral, Kovalenko, Schmidt, and
			Lyubovitskij}}]{Castillo-Felisola:2015ema}
	\bibinfo{author}{\bibfnamefont{O.}~\bibnamefont{Castillo-Felisola}},
	\bibinfo{author}{\bibfnamefont{C.}~\bibnamefont{Corral}},
	\bibinfo{author}{\bibfnamefont{S.}~\bibnamefont{Kovalenko}},
	\bibinfo{author}{\bibfnamefont{I.}~\bibnamefont{Schmidt}}, \bibnamefont{and}
	\bibinfo{author}{\bibfnamefont{V.~E.} \bibnamefont{Lyubovitskij}},
	\bibinfo{journal}{Phys. Rev.} \textbf{\bibinfo{volume}{D91}},
	\bibinfo{pages}{085017} (\bibinfo{year}{2015}).
	
	\bibitem[{\citenamefont{Gutsche et~al.}(2017)\citenamefont{Gutsche,
			Hiller~Blin, Kovalenko, Kuleshov, Lyubovitskij, Vicente~Vacas, and
			Zhevlakov}}]{Gutsche:2016jap}
	\bibinfo{author}{\bibfnamefont{T.}~\bibnamefont{Gutsche}},
	\bibinfo{author}{\bibfnamefont{A.~N.} \bibnamefont{Hiller~Blin}},
	\bibinfo{author}{\bibfnamefont{S.}~\bibnamefont{Kovalenko}},
	\bibinfo{author}{\bibfnamefont{S.}~\bibnamefont{Kuleshov}},
	\bibinfo{author}{\bibfnamefont{V.~E.} \bibnamefont{Lyubovitskij}},
	\bibinfo{author}{\bibfnamefont{M.~J.} \bibnamefont{Vicente~Vacas}},
	\bibnamefont{and}
	\bibinfo{author}{\bibfnamefont{A.}~\bibnamefont{Zhevlakov}},
	\bibinfo{journal}{Phys. Rev.} \textbf{\bibinfo{volume}{D95}},
	\bibinfo{pages}{036022} (\bibinfo{year}{2017}).
	
	\bibitem[{\citenamefont{Zhevlakov et~al.}(2019)\citenamefont{Zhevlakov,
			Gorchtein, Hiller~Blin, Gutsche, and Lyubovitskij}}]{Zhevlakov:2018rwo}
	\bibinfo{author}{\bibfnamefont{A.~S.} \bibnamefont{Zhevlakov}},
	\bibinfo{author}{\bibfnamefont{M.}~\bibnamefont{Gorchtein}},
	\bibinfo{author}{\bibfnamefont{A.~N.} \bibnamefont{Hiller~Blin}},
	\bibinfo{author}{\bibfnamefont{T.}~\bibnamefont{Gutsche}}, \bibnamefont{and}
	\bibinfo{author}{\bibfnamefont{V.~E.} \bibnamefont{Lyubovitskij}},
	\bibinfo{journal}{Phys. Rev.} \textbf{\bibinfo{volume}{D99}},
	\bibinfo{pages}{031703(R)} (\bibinfo{year}{2019}).
	
	\bibitem[{\citenamefont{Tiator et~al.}(2018)\citenamefont{Tiator, Gorchtein,
			Kashevarov, Nikonov, Ostrick, Had\v{z}imehmedovi\'{c}, Omerovi\'{c},
			Osmanovi\'{c}, Stahov, and \v{S}varc}}]{Tiator:2018heh}
	\bibinfo{author}{\bibfnamefont{L.}~\bibnamefont{Tiator}},
	\bibinfo{author}{\bibfnamefont{M.}~\bibnamefont{Gorchtein}},
	\bibinfo{author}{\bibfnamefont{V.~L.} \bibnamefont{Kashevarov}},
	\bibinfo{author}{\bibfnamefont{K.}~\bibnamefont{Nikonov}},
	\bibinfo{author}{\bibfnamefont{M.}~\bibnamefont{Ostrick}},
	\bibinfo{author}{\bibfnamefont{M.}~\bibnamefont{Had\v{z}imehmedovi\'{c}}},
	\bibinfo{author}{\bibfnamefont{R.}~\bibnamefont{Omerovi\'{c}}},
	\bibinfo{author}{\bibfnamefont{H.}~\bibnamefont{Osmanovi\'{c}}},
	\bibinfo{author}{\bibfnamefont{J.}~\bibnamefont{Stahov}}, \bibnamefont{and}
	\bibinfo{author}{\bibfnamefont{A.}~\bibnamefont{\v{S}varc}},
	\bibinfo{journal}{Eur. Phys. J.} \textbf{\bibinfo{volume}{A54}},
	\bibinfo{pages}{210} (\bibinfo{year}{2018}).
	
	\bibitem[{\citenamefont{Weinberg}(1968)}]{Weinberg:1968de}
	\bibinfo{author}{\bibfnamefont{S.}~\bibnamefont{Weinberg}},
	\bibinfo{journal}{Phys. Rev.} \textbf{\bibinfo{volume}{166}},
	\bibinfo{pages}{1568} (\bibinfo{year}{1968}).
	
	\bibitem[{\citenamefont{Faessler
			et~al.}(2006{\natexlab{c}})\citenamefont{Faessler, Gutsche, Lyubovitskij, and
			Pumsa-ard}}]{Faessler:2005gd}
	\bibinfo{author}{\bibfnamefont{A.}~\bibnamefont{Faessler}},
	\bibinfo{author}{\bibfnamefont{T.}~\bibnamefont{Gutsche}},
	\bibinfo{author}{\bibfnamefont{V.~E.} \bibnamefont{Lyubovitskij}},
	\bibnamefont{and}
	\bibinfo{author}{\bibfnamefont{K.}~\bibnamefont{Pumsa-ard}},
	\bibinfo{journal}{Phys. Rev.} \textbf{\bibinfo{volume}{D73}},
	\bibinfo{pages}{114021} (\bibinfo{year}{2006}{\natexlab{c}}).
	
	\bibitem[{\citenamefont{Lensky and Pascalutsa}(2010)}]{Lensky:2009uv}
	\bibinfo{author}{\bibfnamefont{V.}~\bibnamefont{Lensky}} \bibnamefont{and}
	\bibinfo{author}{\bibfnamefont{V.}~\bibnamefont{Pascalutsa}},
	\bibinfo{journal}{Eur. Phys. J.} \textbf{\bibinfo{volume}{C65}},
	\bibinfo{pages}{195} (\bibinfo{year}{2010}).
	
	\bibitem[{\citenamefont{Aaij et~al.}(2017)}]{Aaij:2016jaa}
	\bibinfo{author}{\bibfnamefont{R.}~\bibnamefont{Aaij}} \bibnamefont{et~al.}
	(\bibinfo{collaboration}{LHCb Collaboration}), \bibinfo{journal}{Phys. Lett.}
	\textbf{\bibinfo{volume}{B764}}, \bibinfo{pages}{233} (\bibinfo{year}{2017}).
	
	\bibitem[{\citenamefont{Crewther et~al.}(1979)\citenamefont{Crewther,
			Di~Vecchia, Veneziano, and Witten}}]{Crewther:1979pi}
	\bibinfo{author}{\bibfnamefont{R.~J.} \bibnamefont{Crewther}},
	\bibinfo{author}{\bibfnamefont{P.}~\bibnamefont{Di~Vecchia}},
	\bibinfo{author}{\bibfnamefont{G.}~\bibnamefont{Veneziano}},
	\bibnamefont{and} \bibinfo{author}{\bibfnamefont{E.}~\bibnamefont{Witten}},
	\bibinfo{journal}{Phys. Lett.} \textbf{\bibinfo{volume}{88B}},
	\bibinfo{pages}{123} (\bibinfo{year}{1979}), \bibinfo{note}{{\bf 91B}, 487(E)
		(1980)}.
	
	\bibitem[{\citenamefont{Shifman et~al.}(1980)\citenamefont{Shifman, Vainshtein,
			and Zakharov}}]{ShifmanB166}
	\bibinfo{author}{\bibfnamefont{M.~A.} \bibnamefont{Shifman}},
	\bibinfo{author}{\bibfnamefont{A.~I.} \bibnamefont{Vainshtein}},
	\bibnamefont{and} \bibinfo{author}{\bibfnamefont{V.~I.}
		\bibnamefont{Zakharov}}, \bibinfo{journal}{Nucl. Phys.}
	\textbf{\bibinfo{volume}{B166}}, \bibinfo{pages}{493} (\bibinfo{year}{1980}).
	
	\bibitem[{\citenamefont{Harris et~al.}(1999)}]{Harris:1999jx}
	\bibinfo{author}{\bibfnamefont{P.~G.} \bibnamefont{Harris}}
	\bibnamefont{et~al.}, \bibinfo{journal}{Phys. Rev. Lett.}
	\textbf{\bibinfo{volume}{82}}, \bibinfo{pages}{904} (\bibinfo{year}{1999}).
	
	\bibitem[{\citenamefont{Baker et~al.}(2006)}]{Baker:2006ts}
	\bibinfo{author}{\bibfnamefont{C.~A.} \bibnamefont{Baker}}
	\bibnamefont{et~al.}, \bibinfo{journal}{Phys. Rev. Lett.}
	\textbf{\bibinfo{volume}{97}}, \bibinfo{pages}{131801}
	(\bibinfo{year}{2006}).
	
	\bibitem[{\citenamefont{Gan}(2017)}]{Gan:2017kfr}
	\bibinfo{author}{\bibfnamefont{L.}~\bibnamefont{Gan}}, \bibinfo{journal}{JPS
		Conf. Proc.} \textbf{\bibinfo{volume}{13}}, \bibinfo{pages}{020063}
	(\bibinfo{year}{2017}).
	
	\bibitem[{\citenamefont{Tanabashi et~al.}(2018)}]{Tanabashi:2018oca}
	\bibinfo{author}{\bibfnamefont{M.}~\bibnamefont{Tanabashi}}
	\bibnamefont{et~al.} (\bibinfo{collaboration}{Particle Data Group}),
	\bibinfo{journal}{Phys. Rev.} \textbf{\bibinfo{volume}{D98}},
	\bibinfo{pages}{030001} (\bibinfo{year}{2018}).
	
	\bibitem[{\citenamefont{Gasser and Leutwyler}(1982)}]{Gasser:1982ap}
	\bibinfo{author}{\bibfnamefont{J.}~\bibnamefont{Gasser}} \bibnamefont{and}
	\bibinfo{author}{\bibfnamefont{H.}~\bibnamefont{Leutwyler}},
	\bibinfo{journal}{Phys. Rept.} \textbf{\bibinfo{volume}{87}},
	\bibinfo{pages}{77} (\bibinfo{year}{1982}).
	
	\bibitem[{\citenamefont{Graner et~al.}(2016)\citenamefont{Graner, Chen,
			Lindahl, and Heckel}}]{Graner:2016ses}
	\bibinfo{author}{\bibfnamefont{B.}~\bibnamefont{Graner}},
	\bibinfo{author}{\bibfnamefont{Y.}~\bibnamefont{Chen}},
	\bibinfo{author}{\bibfnamefont{E.~G.} \bibnamefont{Lindahl}},
	\bibnamefont{and} \bibinfo{author}{\bibfnamefont{B.~R.}
		\bibnamefont{Heckel}}, \bibinfo{journal}{Phys. Rev. Lett.}
	\textbf{\bibinfo{volume}{116}}, \bibinfo{pages}{161601}
	(\bibinfo{year}{2016}), \bibinfo{note}{{\bf 119}, 119901(E) (2017)}.
	
	\bibitem[{\citenamefont{Sahoo}(2017)}]{Sahoo:2016zvr}
	\bibinfo{author}{\bibfnamefont{B.}~\bibnamefont{Sahoo}},
	\bibinfo{journal}{Phys. Rev.} \textbf{\bibinfo{volume}{D95}},
	\bibinfo{pages}{013002} (\bibinfo{year}{2017}).
	
\end{thebibliography}
\end{document}